\documentclass[10pt,twocolumn,letterpaper]{article}

\usepackage[pagenumbers]{cvpr} 

\usepackage{times}
\usepackage{epsfig}
\usepackage{graphicx}
\usepackage{amsmath}
\usepackage{amssymb}
\usepackage{booktabs}
\usepackage{color}
\usepackage{multirow}
\usepackage{bbding}
\usepackage[accsupp]{axessibility}
\usepackage[pagebackref,breaklinks,colorlinks]{hyperref}

\usepackage[capitalize]{cleveref}
\crefname{section}{Sec.}{Secs.}
\Crefname{section}{Section}{Sections}
\Crefname{table}{Table}{Tables}
\crefname{table}{Tab.}{Tabs.}

\def\cvprPaperID{84} 
\def\confName{CVPR}
\def\confYear{2022}

\begin{document}

\title{Fast and Memory-Efficient Network Towards Efficient Image Super-Resolution}

\author{Zongcai Du$^{1}$, Ding Liu$^{2}$, Jie Liu$^{1}$, Jie Tang$^{1}$, Gangshan Wu$^{1}$, Lean Fu$^{2}$\\
$^{1}$State Key Laboratory for Novel Software Technology, Nanjing University, China\\
$^{2}$ByteDance Inc.\\
    {\tt\small \{151220022, jieliu\}@smail.nju.edu.cn, \{tangjie, gswu\}@nju.edu.cn,}\\
    {\tt\small liudingdavy@gmail.com, fulean@bytedance.com}
}

\maketitle

\begin{abstract}
   Runtime and memory consumption are two important aspects for efficient image super-resolution (EISR) models to be deployed on resource-constrained devices. 
   Recent advances in EISR\cite{IMDN,RFDN} exploit distillation and aggregation strategies with plenty of channel split and concatenation operations to fully use limited hierarchical features.
   In contrast, sequential network operations avoid frequently accessing preceding states and extra nodes, and thus are beneficial to reducing the memory consumption and runtime overhead.
   Following this idea, we design our lightweight network backbone by mainly stacking multiple highly optimized convolution and activation layers and decreasing the usage of feature fusion.
   We propose a novel sequential attention branch, where every pixel is assigned an important factor according to local and global contexts, to enhance high-frequency details.
   In addition, we tailor the residual block for EISR and propose an enhanced residual block (ERB) to further accelerate the network inference. 
   Finally, combining all the above techniques, we construct a fast and memory-efficient network (FMEN) and its small version FMEN-S, which runs 33\% faster and reduces 74\% memory consumption compared with the state-of-the-art EISR model: E-RFDN, the champion in \cite{zhang2020aim}. Besides, FMEN-S achieves the lowest memory consumption and the second shortest runtime in NTIRE 2022 challenge on efficient super-resolution~\cite{aim2022}. Code is available at \url{https://github.com/NJU-Jet/FMEN}.
\end{abstract}

\begin{figure}[t]
	\centering
	\includegraphics[width=1.0\linewidth]{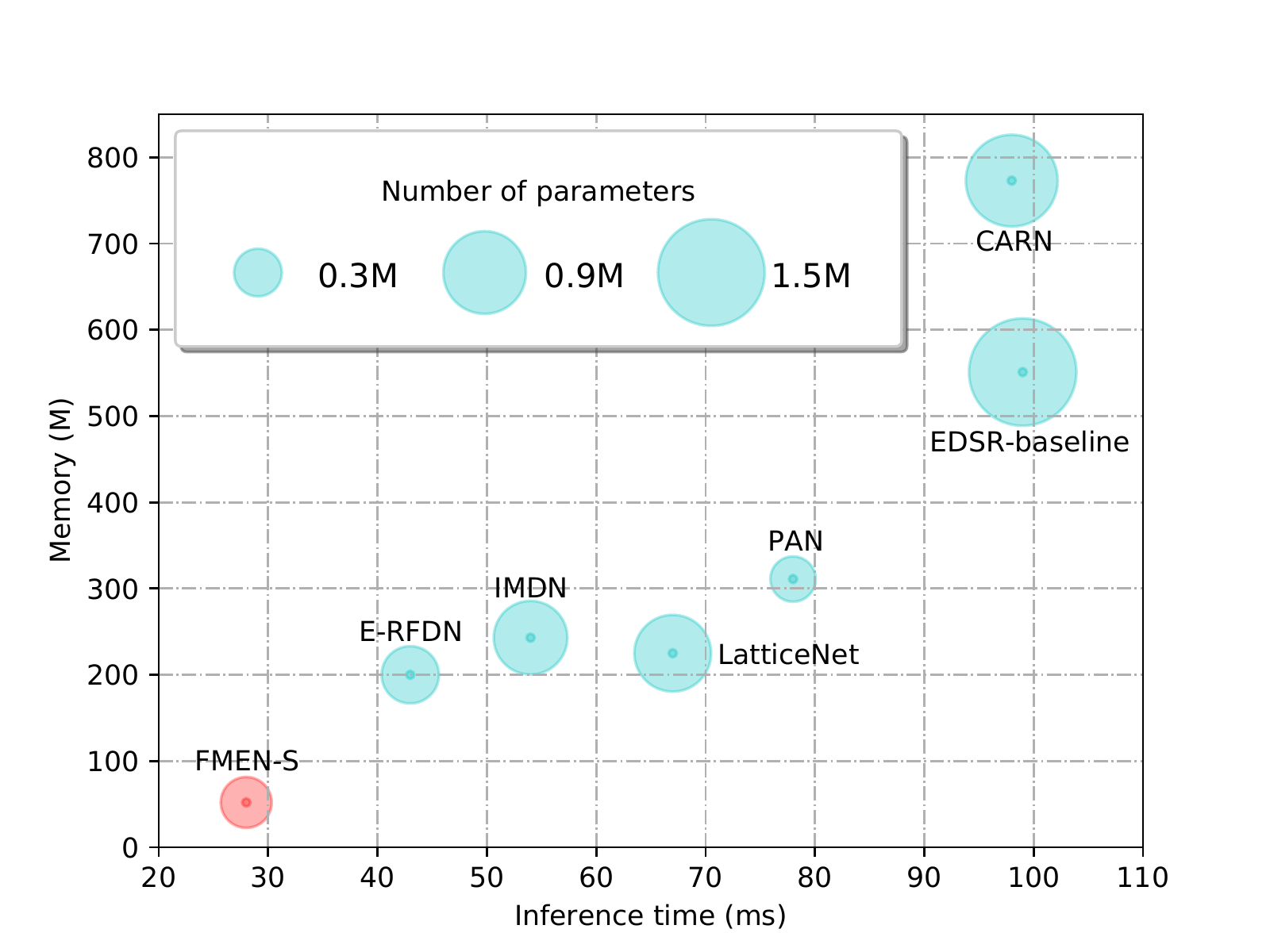} 
	\caption{Comparison with recent EISR models. We show three metrics: inference time on an NVIDIA 1080Ti GPU, memory consumption of input of size $256\times 256$ and the number of parameters for $\times$4 upscaling, when all these models obtain comparable performance on DIV2K validation set (maintain the PSNR of 29.00dB).}
	\label{fig:time}
\end{figure}

\section{Introduction}
Single image super-resolution (SR) is a typical low-level vision problem, with the purpose of recovering a high-resolution (HR) image according to its degraded low-resolution (LR) counterpart. To solve this highly ill-posed problem, different kinds of methods have been proposed. Among them, deep learning based methods~\cite{SRCNN, EDSR, RCAN, IMDN, RFDN}, represented by convolution neural network (CNN), have produced superior results and revolutionized this area.

Due to the realistic demand of resource-limited devices, efficient image super-resolution (EISR) attracts growing attention of the SR community. In the early stage, recurrent neural network~\cite{MemNet, SRFBN} and group convolution~\cite{CARN} are adopted to reduce the model parameters, but they bring about either high computation cost or inferior performance. Recently, many researchers concentrate on elaborate network design, since it not only exerts a direct influence on the performance but also affects the follow-up network pruning~\cite{pruning1, pruning2, DHP, THE} and knowledge distillation~\cite{distillation1, distillation2}. The key point of efficient network design is how to sufficiently utilize limited features to generate more representative features. 

We observe that recent EISR methods tend to apply feature fusion to achieve this goal~\cite{CARN, IMDN, IDN, LatticeNet, RFDN} due to the following two main advantages. First, multi-level connections are conductive to propagating gradients, which facilitates the training of deeper networks. 
Second, it is often associated with the gating mechanism which adaptively controls the previous and current states in feature domain. CARN~\cite{CARN} used cascading mechanism at both local and global levels to incorporate the features from multiple layers. IMDN~\cite{IMDN} proposed information multi-distillation blocks (IMDB) to extract hierarchical features step-by-step and aggregate them according to the importance of candidate features. LatticeNet~\cite{LatticeNet} adopted a backward sequential concatenation strategy for feature fusion of different receptive fields. RFDN~\cite{RFDN} applied  residual feature distillation block which is a variant of IMDB but more powerful and flexible. The feature fusion strategies mentioned above suffer from huge memory consumption that stem from multiple relevant feature maps residing in the memory until aggregated. In addition, the fusion design usually reduces inference speed because of introducing extra nodes (\eg, concatenation, $1\times 1$ convolution) and frequent memory access as discussed in RepVGG~\cite{RepVGG}.
To accelerate inference speed and reduce memory consumption, we design our network backbone via highly optimized serial network operations instead of feature aggregation, and boost feature representations via attention mechanism. 

Considering the goal of SR is to recover the lost high-frequency details (\eg, edges, textures), we propose a high-frequency attention block (HFAB) which learns an attention map with special focus on the high-frequency area.
Specifically, we design the attention branch in HFAB from local and global perspectives. We stack highly efficient operators like $3\times 3$ convolution and Leaky ReLU layers sequentially for modeling the relationship between local signals. Batch Normalization (BN) is injected into HFAB to capture global context during training, while merged into convolution during inference. 
Furthermore, we tailor the residual block (RB) and introduce an enhanced residual block (ERB), where the features are extracted in higher dimensional space during training and the skip connections are removed during inference using structural re-parameterization technique~\cite{RepVGG}. We show that such a design is able to accelerate the network inference and reduce memory consumption without sacrificing SR performance, when comparing with conventional RB. 

By applying ERB and HFAB in a sequential and alternative way, we construct an efficient network, namely fast and memory-efficient network (FMEN), which demonstrates the clear advantage over existing EISR methods in terms of runtime and peak memory consumption when maintaining the same level of restoration performance. Besides, we build a smaller model by reducing the number of convolution kernels, FMEN-S, which runs 33\% faster and reduces 74\% memory consumption compared with E-RFDN as shown in Fig.~\ref{fig:time}, our contributions are as follows:
\begin{itemize}
    \item We carefully analyze the factors which influence the inference speed and memory consumption of EISR models.
    \item We propose a high-frequency attention block (HFAB) to enchance high-frequency features and an enhanced residual block (ERB) to utilize residual learning with faster inference and less memory consumption. 
    \item We construct a fast and memory-efficient network by sequentially combining HFAB and ERB, which achieves the lowest memory consumption and the second shortest runtime in NTIRE 2022 challenge on efficient super-resolution.
\end{itemize}

\section{Related Work}
\subsection{Overview of Image Super-Resolution}
Recently, convolution neural network based (CNN-based) methods have attained excellent results in many tasks including SISR. Since Dong \etal~\cite{SRCNN} creatively introduced a three-layer end-to-end CNN called SRCNN to restore HR image, diverse methods have been proposed to further enhance learning capability. To reduce its high computational cost which is caused by learning in HR space, shi \etal designed ESPCN~\cite{ESPCN} to replace the bicubic filter with sub-pixel convolution, which is adopted by mainstream SR architectures~\cite{ESRGAN, RFDN, LatticeNet, ECBSR, BasicVSR}. In the same period, Kim \etal~\cite{VDSR} used residual learning and adjustable gradient clipping to train an extremely accurate twenty-layer model, demonstrating that depth plays an important role in the reconstruction performance. Subsequently, Ledig \etal~\cite{SRResNet} successfully exploited residual block (RB)~\cite{ResNet} and presented a GAN-based network to recover photo-realistic textures. Furthermore, based on SRResNet~\cite{SRResNet}, Lim \etal~\cite{EDSR} proposed an enhanced deep super-resolution network (EDSR), which surpassed previous networks by removing unnecessary modules in RB and inspired succeeding works~\cite{RDN, RCAN, CARN, RNAN, RFA, LatticeNet}. For instance, RDN~\cite{RDN} presented residual dense block to make full use of all the hierarchical features via dense connected convolution layers. RCAN~\cite{RCAN} integrated channel attention mechanism into RB and adopted residual-in-residual (RIR) structure to form a very deep network. LatticeNet~\cite{LatticeNet} applied butterfly structure to adaptively combine two RBs. 

\subsection{Efficient Image Super-Resolution}
In order to deploy SR models, several aspects should be also taken into consideration except restoration accuracy, such as the number of parameters, FLOPs, peak memory consumption and inference time. Recent methods towards EISR can be roughly divided into explicit~\cite{DRCN, DRRN, SRFBN, CARN, IDN} and implicit schemes~\cite{LapSRN, CARN, MemNet, IMDN, LatticeNet}. The former reduces the model complexity via cutting down the width and depth~\cite{SRFBN}, recurrent structure~\cite{DRCN, DRRN} and group convolution~\cite{CARN, IDN}. The aforementioned strategies cause either accuracy loss or more extra overheads (\eg, FLOPs). The latter implicit scheme focuses on sufficiently leveraging middle features as well as enhancing the discriminative ability, thus leading to lower complexity and better performance on the whole. For example, LapSRN~\cite{LapSRN} took advantage of layered pyramid features to reconstruct residuals of 
various resolution. MemNet~\cite{MemNet} adopted gating mechanism to bridge deep features with shallow information. CARN~\cite{CARN} presented local and global cascading mechanism inspired by SRDenseNet~\cite{SRDenseNet} to boost up the representation power. IMDN~\cite{IMDN} retained partial features as refined information and aggregated the distilled features via contrast-aware channel attention block. LatticeNet~\cite{LatticeNet} applied a butterfly structure to dynamically combine two RBs. RFDN~\cite{RFDN} enhances IMDB via feature distillation connection. DLSR~\cite{DLSR} proposes a differentiable neural architecture search approach to find more powerful fusion blocks based on RFDB. It can be seen that feature fusion plays a pivotal role in recent advances. While theoretically efficient, it is sub-optimal due to relevant features residing in the memory until aggregation, leading to multiple times of memory consumption compared with simple topology counterparts without feature fusion.

\subsection{Attention Mechanism in SR}
Attention mechanism has been shown to be extremely powerful and applied in conjunction with CNN for various computer vision tasks, including SR~\cite{SENet, CBAM, RCAN, RNAN, SAN, IMDN, RFDN}. It aims at guiding the network to focus on important signals and suppress unnecessary ones. Since the success of SENet~\cite{SENet} for image classification, various kinds of attention mechanisms have been applied to SR models. RCAN~\cite{RCAN} first integrated channel attention (CA) into RB for SISR. RNAN~\cite{RNAN} introduced local and non-local blocks to adaptively rescale the hierarchical features. SAN~\cite{SAN} used second-order feature statistics to generate more representative channel attention map. RFA~\cite{RFA} employed an enhanced spatial attention (ESA) block to obtain more sophisticated attention map. HAN~\cite{HAN} proposed layer attention and channel-spatial attention to model the holistic inter-dependencies among layers, channels, and positions. Recently, transformer-based methods~\cite{IPT, SwinIR} have also been introduced to SR. SwinIR~\cite{SwinIR} applied shifted window mechanism to model long-range dependency, which can be interpreted as spatially varying convolution for capturing content-based interactions between image content and attention weights. They have achieved significant progress, but are not efficient enough for EISR due to multi-branch topology, multiple features residing in the memory and inefficient operations.

\begin{figure}[t]
	\centering
	\includegraphics[width=1.0\linewidth]{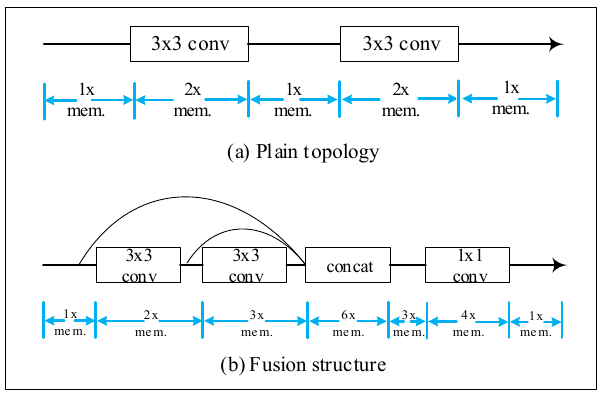} 
	\caption{Peak memory consumption of plain topology and fusion structure during model inference. Suppose feature size remains the same after $3\times 3$ convolution. Multiple features are fused in the concatenation node along the channel dimension and $1\times 1$ convolution is used to reduce the final output feature size to the initial input feature size in (b). Activation layers are omitted for simplicity. 
	}
	\label{fig:memory}
\end{figure}

\begin{figure*}[htbp]
	\centering
	\includegraphics[width=1.0\linewidth]{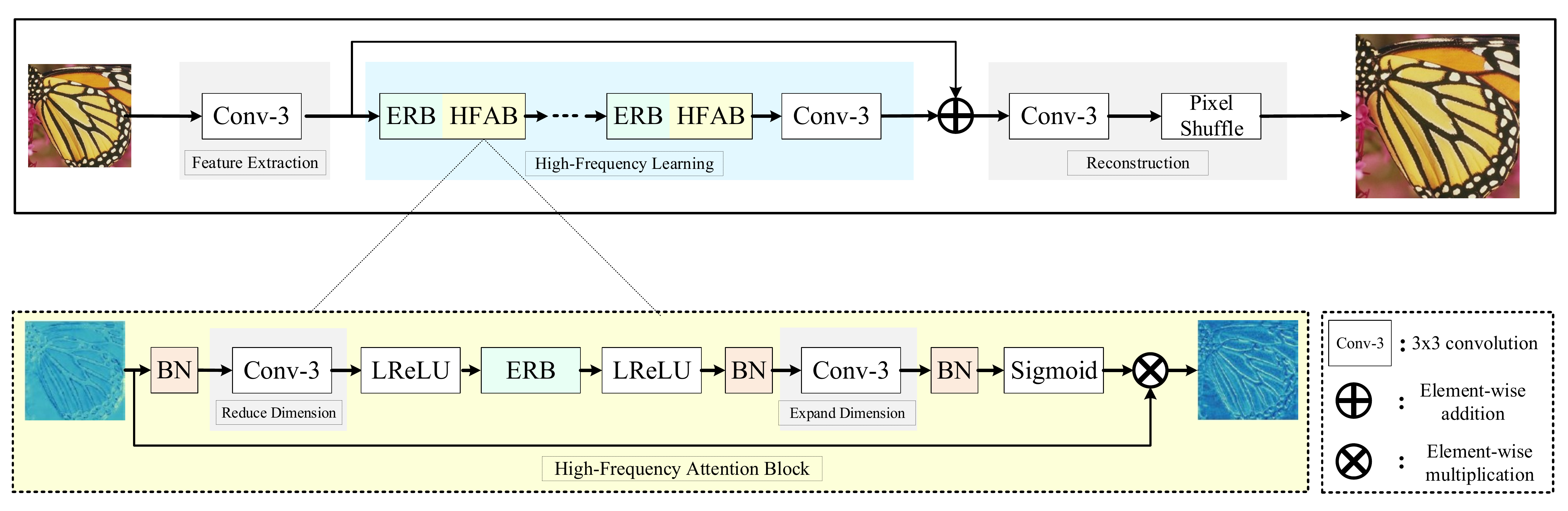} 
	\caption{The overall architecture of fast and memory-efficient network (FMEN).}
	\label{fig:architecture}
\end{figure*}
\section{Proposed Method}
\subsection{Memory Analysis}
\label{sec:memory}
We first introduce the memory analysis during inference which motivates the design of our network architecture.
Memory consumption is an important factor for EISR model deployment. Generally, the memory consumption $M$ at one node is composed of four parts: input feature memory $M_{input}$, output feature memory $M_{output}$, kept feature memory $M_{kept}$ which is calculated before and used in the future nodes, network parameter memory $M_{net}$. The total memory $M$ can be formulated as:
$M = M_{input} + M_{output} + M_{kept} + M_{net}$.
$M_{net}$ is too small and thus negligible compared with feature memory. 
To compare the memory consumption of plain network topology and feature fusion schemes,
we consider two typical structures of plain topology and feature fusion of EISR in Fig.~\ref{fig:memory} (suppose the input occupies 1x memory). 
For inplace ReLU layer, the input and output share the same memory chunk so $M_{output}$ is zero, and we omit it for the simplicity of discussion.
For a convolution layer with kernel size $C_{in}\times C_{out}\times K\times K$, $M_{input}$ and $M_{output}$ cannot be shared due to $C_{out}\times K\times K$ visits of every input location and Winograd~\cite{Winograd}. 

Considering plain topology of a common $3\times 3$ convolution in Fig.~\ref{fig:memory}(a), the peak memory consumption in the convolution node is determined by $M_{input}+M_{output}$. If the feature size is not changed in the whole process, the peak memory consumption for this plain topology is around $2\times C\times H\times W$, where $C$ is the number of feature maps and $H$, $W$ denote the height and width of each feature map, respectively.
Stacking the same network topology sequentially does not increase the peak memory consumption during inference.

As for the fusion structure, feature maps related to the fusion layer need to be kept until concatenation is finished.  
Taking the case of Fig.~\ref{fig:memory}(b) as an example, there are three features occupying memory in the second $3\times 3$ convolution layer: the input and output of this layer, the input of the first $3\times 3$ convolution layer which will be used in the latter concatenation node, so the peak memory consumption for this layer is $3\times$ as the input. By the same logic, peak memory consumption doubles at the following concatenation node of three input features.
In general, if there are $N$ features of the same size ($C\times H\times W)$ taking part in fusion, the memory consumption is raised at least to $2\times N\times C\times H\times W$, since $M_{input}$ and $M_{output}$ are both $N\times C\times H\times W$ for the concatenation node. 
If global and local fusions are used simultaneously like in current lightweight architectures~\cite{CARN, MemNet, IMDN, RFDN}, $N_{local}$ input features, $N_{local}$ concatenated output features and $N_{global}-1$ kept features will occupy memory concurrently at the concatenation node of the last local fusion part, where $N_{global}$ and $N_{local}$ are the number of features taking part in the global and local fusions, respectively.
%
Therefore, feature fusion generally increases peak memory consumption during inference in comparison with plain sequential topology.

\subsection{Network Architecture}
Applying sequential network topology to EISR is not a trivial task.
One way is to directly adopt fully sequential architecture~\cite{FSRCNN},
%
and another way is to replace normal convolution layers with re-parameterizable building blocks~\cite{ECBSR} to expand optimization space during training. However, both often suffer the performance drop comparing with recent advanced fusion topology~\cite{LatticeNet, RFDN}. 
%

In addition to memory consumption, inference time is another key aspect for EISR models. 
Based on the sequential network topology, 
we propose an enhanced residual block (ERB) for deep feature learning and an effective high-frequency attention block (HFAB) for feature enhancement, both of which not only reduce memory consumption but also accelerate the inference. The overall network architecture is shown in Fig.~\ref{fig:architecture}, which contains three main parts: feature extraction part, detail learning part and reconstruction part. The first and the last parts are kept the same as previous works~\cite{EDSR, IMDN, LatticeNet, ECBSR}. The second part is composed of alternating ERB and HFAB. The peak memory consumption of the entire model is reached at each element-wise multiplication node inside HFAB, where the memory is occupied by global residual, input feature of corresponding HFAB, the attention map and the output feature of corresponding HFAB, and is about $4\times C\times H\times W$.

\begin{figure}[h]
	\centering
	\includegraphics[width=1.0\linewidth]{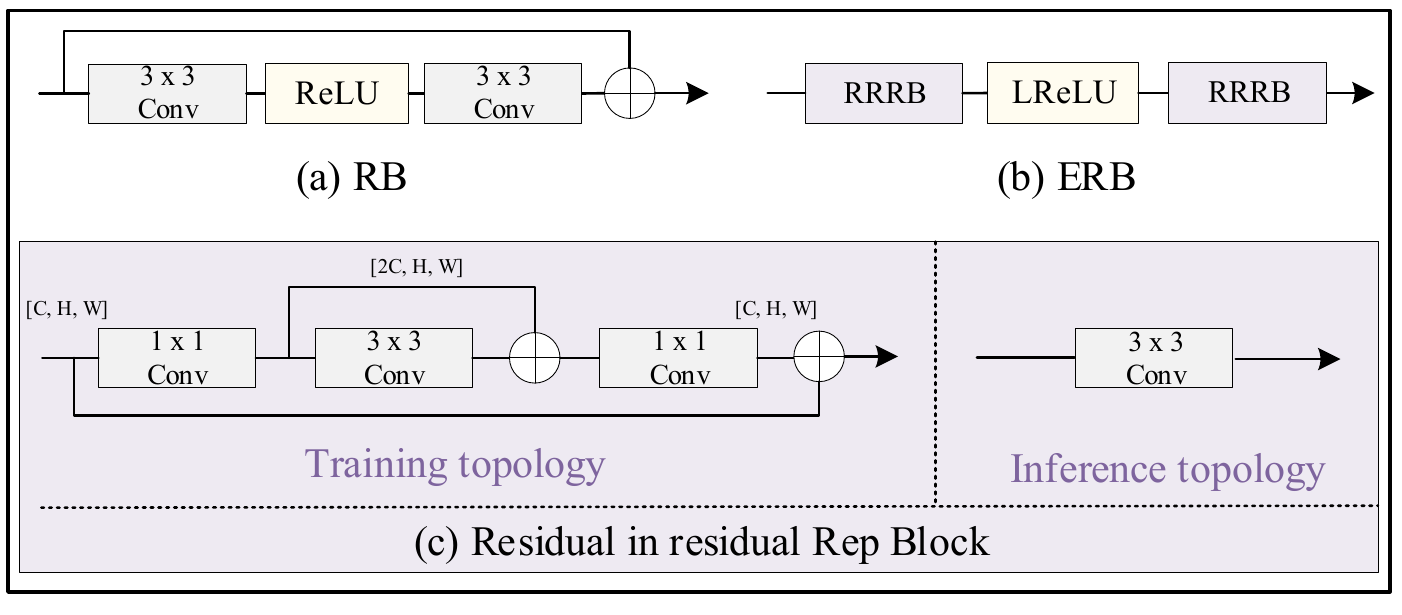} 
	\caption{Architecture of vanilla RB proposed in EDSR~\cite{EDSR} and our proposed ERB.}
	\label{fig:ERB}
\end{figure}
\subsection{Enhanced Residual Block}
Residual block (RB) proposed in \cite{ResNet} has been widely used in network models. It was customized and shown effective for SR in EDSR~\cite{EDSR} and thus has been used in plenty of succeeding SR studies~\cite{CARN, RCAN, RNAN, LatticeNet}. We refer to RB as the version from EDSR in the following unless otherwise stated.
However, the use of skip connection introduces extra memory consumption of feature map size $C\times H\times W$, and lowers the inference speed due to additional memory access cost, which is shown experimentally that if the skip connection is removed in EDSR-baseline~\cite{EDSR} the model runtime is reduced about 10\% in Sec.~\ref{ex:ERB}.
To inherit the merit of residual learning without introducing the aforementioned cost, we design ERB to replace RB. ERB is composed of one Leaky ReLU non-linearity and two residual in residual re-parameterization blocks (RRRB), which is inspired by RCAN~\cite{RCAN} and RepVGG~\cite{RepVGG}. RRRB excavates the potential ability of sophisticated structure during optimization, while being equivalent to a single $3\times3$ convolution during inference. The network structure comparison between RB and ERB is illustrated in Fig.~\ref{fig:ERB}.

\subsection{High-Frequency Attention Block}
Recently, attention mechanism has been extensively studied in the SR literature. Based on the grain-size composition, it can be divided into channel attention~\cite{SENet, IMDN}, spatial attention~\cite{CBAM, RFA}, pixel attention~\cite{PAN}, and layer attention~\cite{HAN}.
Previous attention blocks~\cite{IMDN, RFDN, HAN} are multi-branch topology and contain inefficient operators, which cause extra memory consumption as discussed in Sec.~\ref{sec:memory} and slow the inference speed.
Considering both aspects, we design a high-frequency attention block (HFAB) as shown in Fig.~\ref{fig:architecture}. The attention branch is responsible for assigning a scaling factor to every pixel, and high-frequency areas are expected to be assigned larger values since they mainly influence the restoration accuracy~\cite{SFP, PISR}. We first reduce the channel dimension for efficiency by $3\times$3 convolution instead of $1\times1$ convolution. Afterwards ERB is applied to capture local interaction. Next, channel dimension increases to the original level and a sigmoid layer is used to restrict the value from 0 to 1. Finally, input features are recalibrated by multiplying the attention map in a pixel-wise manner. The motivation of the above steps is mainly from edge detection, where the linear combination of nearby pixels can be used to detect edges. The receptive field brought by convolutions is very limited, which means only the local-range dependency is modeled to determine the importance of every pixel. Thus, batch normalization (BN) is injected into the sequential layers to introduce global interaction, while being beneficial to the unsaturated area of the sigmoid function. Although previous works~\cite{EDSR, ESRGAN} reported that BN could lead to some unexpected artifacts, we empirically observe that BN plays a role in restricting the pixel range diversity, which is in line with the high-frequency learning design, thus contributing to the performance improvement. During inference, we remove the skip connection of ERB inside HFAB and BN layers by merging the corresponding parameters into related convolution layers. HFAB only contains four highly optimized operators: $3\times 3$ convolution, Leaky ReLU non-linearity, sigmoid and element-wise multiplication. HFAB avoids complex multi-branch topology, and thus ensures faster inference in comparison with ESA~\cite{RFA}. Rescaling features in the pixel level makes HFAB more powerful than CCA~\cite{IMDN} in the channel level. The experiment details can be found in Sec.~\ref{ex:attention}.
	\begin{figure}[t]
		\begin{center}
			\begin{tabular}{@{}c@{}c@{}}
                \includegraphics[width=0.5\linewidth]{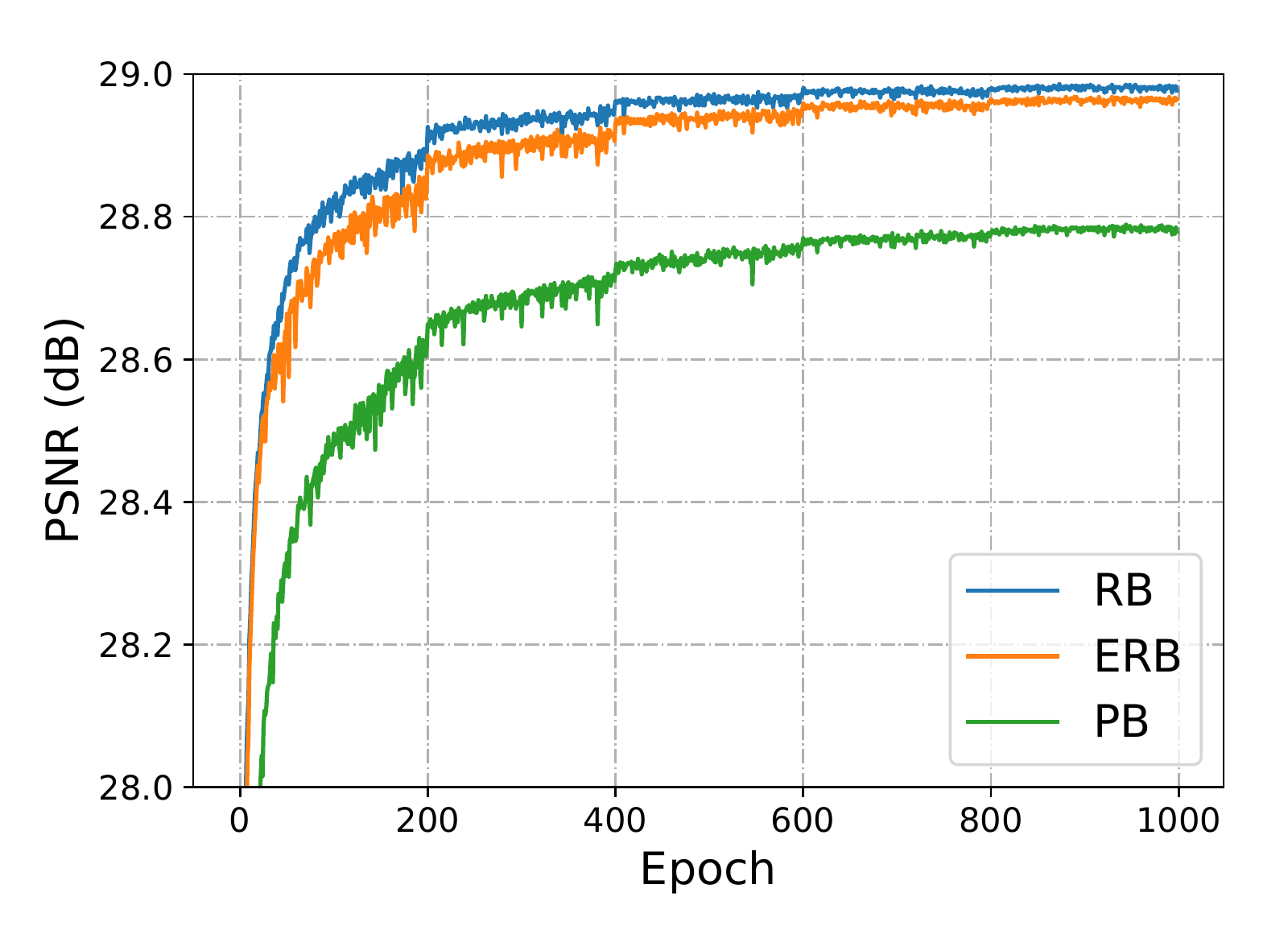} & 
                \includegraphics[width=0.5\linewidth]{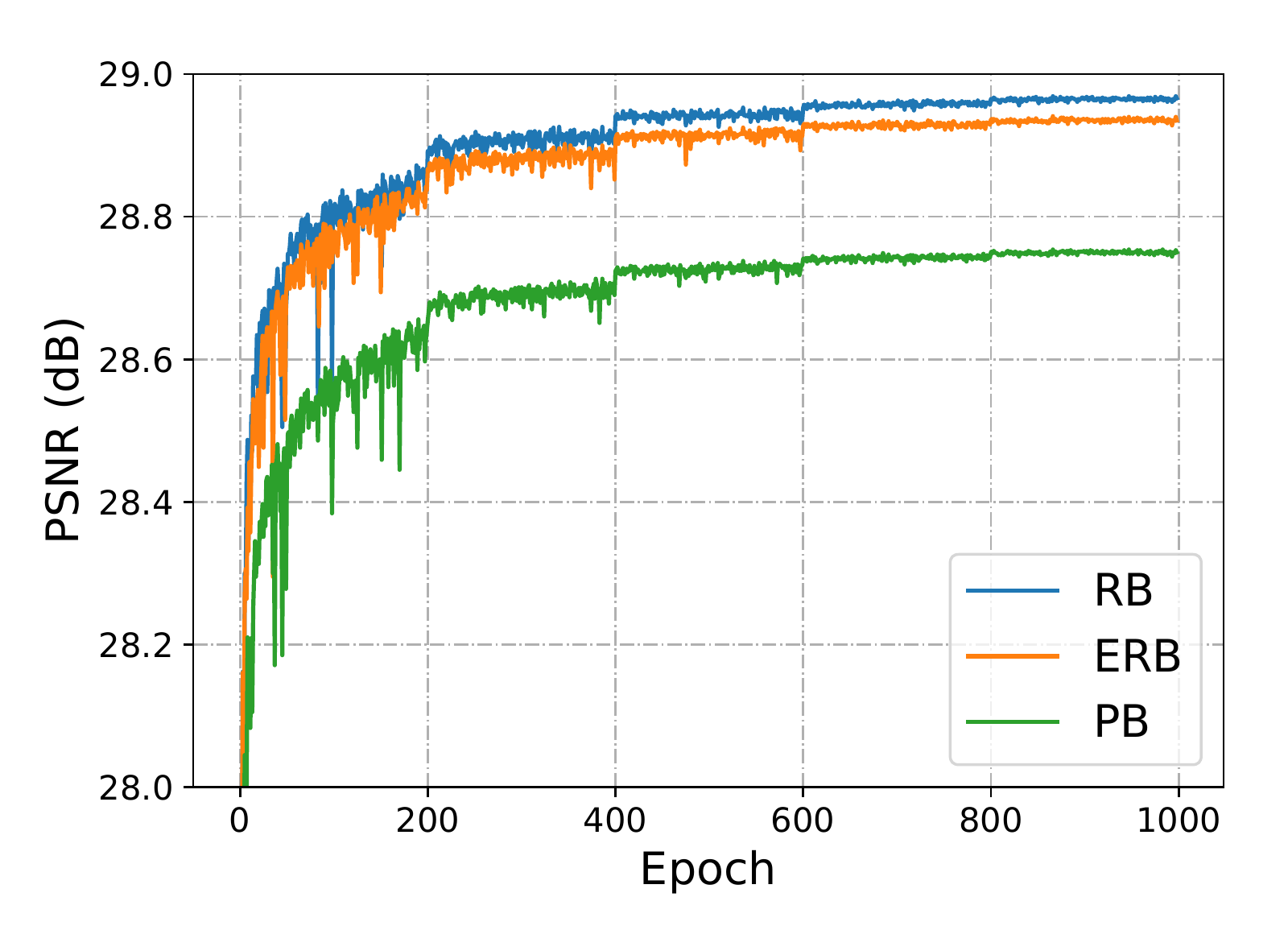}\\
                \quad(a) EDSR-baseline~\cite{EDSR} & \quad(b) FMEN
			\end{tabular}
		\end{center}
	    \caption{PSNR comparison of three blocks by applying them to EDSR-baseline~\cite{EDSR} and FMEN: PB (remove skip connection in RB), RB~\cite{EDSR} and proposed ERB. The performance is reported on DIV2K validation Set for x4 upscaling.
        }
		\label{fig:ablationERB}	
	\end{figure}

\begin{figure*}[htp]
	\centering
	\includegraphics[width=1.0\linewidth]{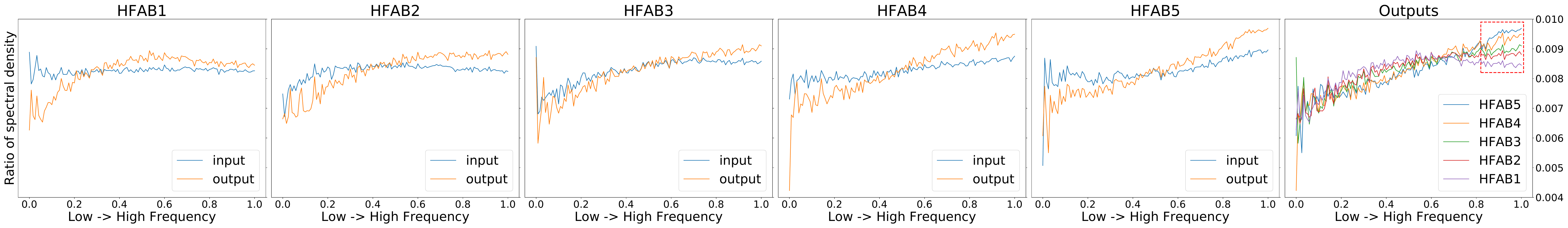} 
	\caption{Frequency analysis of input and output in each HFAB.}
	\label{fig:frequency}
\end{figure*}

\begin{figure}[htp]
	\centering
	\includegraphics[width=1.0\linewidth]{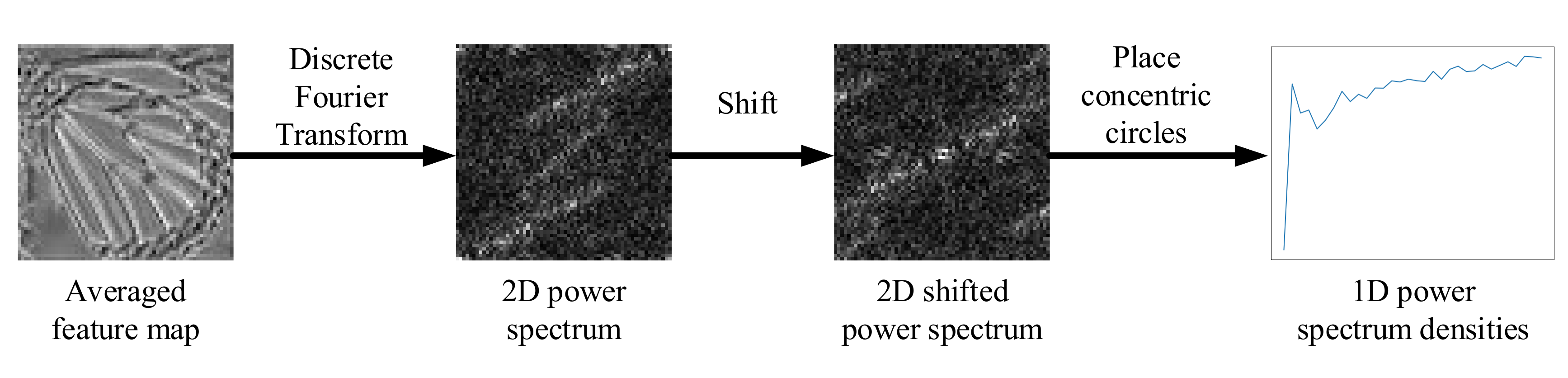} 
	\caption{Frequency analysis pipeline~\cite{MemNet, SRFBN}.}
	\label{fig:psd}
\end{figure}

\begin{figure}[htbp]
	\begin{center}
		\begin{tabular}{@{}c@{}}
			\includegraphics[width=1.0\linewidth]{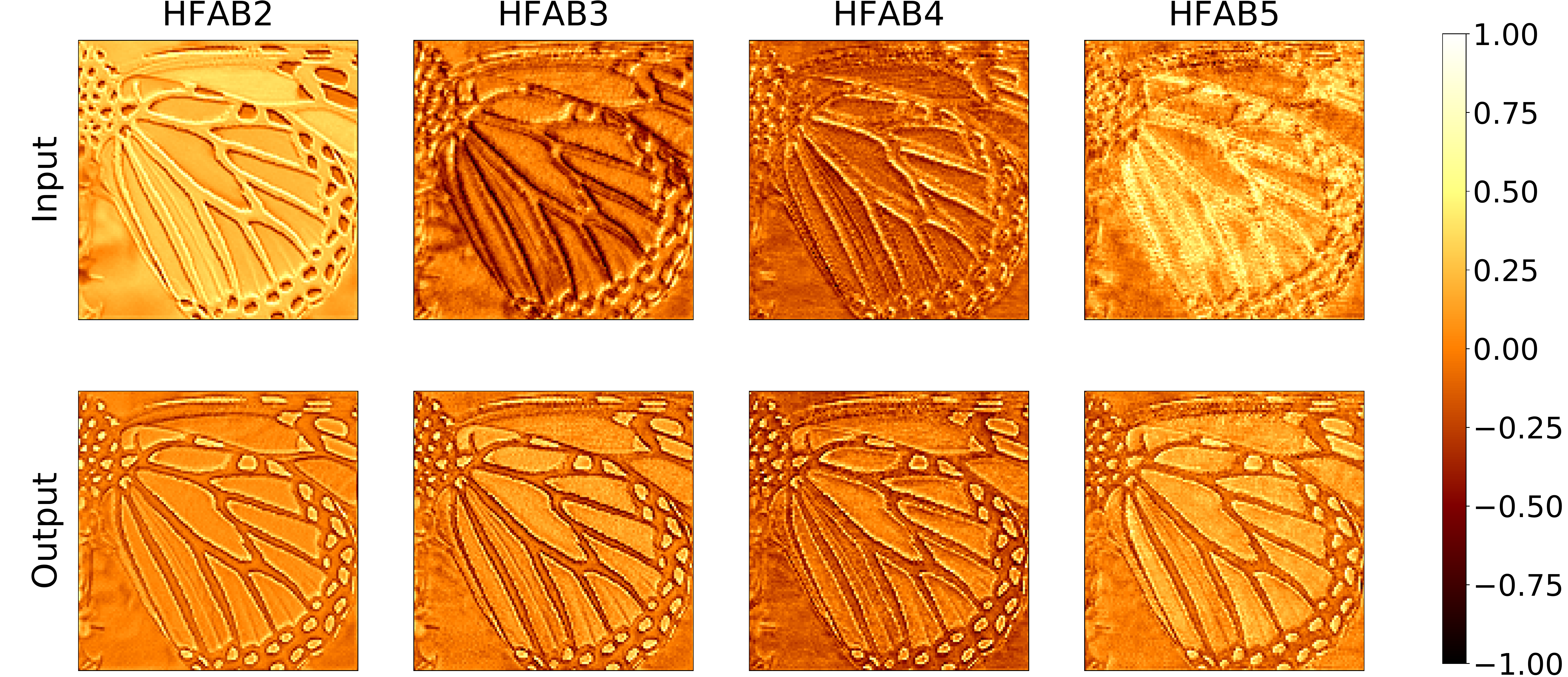}\\
		\end{tabular}
	\end{center}
	\caption{Average feature maps of the four fine grained HFABs. Since the absolute values of outputs are always smaller than the inputs', we normalize them for better visualization.}
	\label{fig:feature}
\end{figure}


\section{Experiments}
\subsection{Settings}
\textbf{Datasets and metrics.}
Following recent works~\cite{RFDN, LatticeNet}, we adopt widely used high-quality (2K resolution) DIV2K~\cite{DIV2K} and Flickr2K as training dataset. For testing, five standard benchmark datasets are used: Set5~\cite{Set5}, Set14~\cite{Set14}, B100~\cite{B100}, Urban100~\cite{Urban100}, and Manga109~\cite{Manga109}. The LR images are generated in the same way as~\cite{EDSR, CARN, IMDN, LatticeNet}. To keep consistence with existing methods~\cite{EDSR, IDN, MemNet, IMDN, LatticeNet}, the SR results are evaluated on the luminance channel of transformed YCbCr space, using two common metrics called peak signal-to-noise ratio (PSNR) and structure similarity index (SSIM)~\cite{PSNR}. Following AIM 2020~\cite{zhang2020aim}, maximum GPU memory consumption is tested using PyTorch function $torch.cuda.max\_memory\_allocated$ on the LR image of size $256\times 256$.

\textbf{Implementation details.} In each training batch, 64 cropped $64 \times 64$ LR RGB patches augmented by random flipping and rotation are input to the network. L1 loss is adopted to train our model. The learning rate is initialized as $5\times 10^{-4}$ and decreases half per $6\times10^5$ iterations for total $3\times10^6$ iterations. The number of convolution kernels in the trunk of ERB is set to 64 and 64, and the first convolution in HFAB reduces the channel dimension to 32. The number of ERB and HFAB pairs is set to 5 to achieve comparable performance as LatticeNet~\cite{LatticeNet}. Parameters of our model are initialized using the method proposed by He \etal~\cite{kaiming} and optimized by ADAM optimizer~\cite{ADAM} with $\beta_1=0.9$, $\beta_2=0.999$, and $\epsilon=10^{-8}$. Our model is implemented using the PyTorch framework on an Nvidia 1080Ti GPU.

\begin{table}[!htbp]
	\centering
	\caption{Inference time comparison between RB and ERB. It is the average of 10 runs on DIV2K validation set for x4 upscaling.}
	\label{tab:time}
	\resizebox{1.0\linewidth}{!}{
		\begin{tabular}{|c|c|c|c|}
		\hline
		& MemNet~\cite{MemNet} & EDSR-baseline~\cite{EDSR} & FMEN\\
		\hline
		RB & 231.7ms & 99.8ms & 46.1ms\\
		\hline
		ERB & 214.5ms ($\downarrow$7.5\%) & 90.6ms ($\downarrow$9.2\%) & 40.4ms ($\downarrow$12.4\%)\\
		\hline
	    \end{tabular}}
\end{table}

\subsection{Effectiveness of ERB}
\label{ex:ERB}
ERB is proposed to combine the advantages of sequential topology and residual learning. 
To validate its effectiveness, we compare three blocks: plain block (PB) by removing the skip connection in RB, RB and ERB. They are of the same number of parameters and we apply them separately to EDSR-baseline~\cite{EDSR} and FMEN. In detail, original EDSR-baseline~\cite{EDSR} adopts RBs to construct the network, we replace it with PBs or ERBs. For FMEN, we replace ERBs with RBs or PBs. From Fig.~\ref{fig:ablationERB}, we can see that ERB architecture achieves comparable performance as RB architecture, while PB architecture behaves much worse (nearly 0.2dB drop). During inference, ERB can be reducible to PB thus enjoys the efficiency of sequential topology.
We also report the inference speed comparison in Tab.~\ref{tab:time}. ERB decreases around 10\% inference time by avoiding memory access cost (MAC).
%
\begin{table}[htbp]
	\centering
	\caption{Investigation of different attention mechanisms. All models are validated on Set5 and Urban100 for $\times4$ upscaling in 300 epochs. We also report the average inference time on DIV2K validation set for $\times4$ upscaling on an Nvidia 1080Ti GPU. The last column shows time increase rate compared with Baseline.}
	\label{tab:attention}
	\resizebox{1.0\linewidth}{!}{
		\begin{tabular}{|c|c|c|c||c|c|}
			\hline
			Model & Params & Set5 & Urban100 & Inference time & Increase rate \\
			\hline
			\hline
			Baseline & 779K & 32.03 & 25.89 & 45.25ms & 0.0$\%$ \\
			Baseline+CCA~\cite{IMDN} & 782K & 32.11 & 25.93 & 50.55ms & 11.7$\%$ \\
			Baseline+ESA~\cite{RFDN} & 784K & 32.12 & 25.95 & 52.11ms & 15.2$\%$ \\
			\hline
			\hline
			FMEN & 769K & \textbf{32.17} & \textbf{26.04} & 46.13ms & 1.9$\%$ \\
			\hline			
	\end{tabular}}
\end{table}
\begin{table*}[t]
	\centering
	\caption{Quantitative results on benchmark datasets. {\color{red}Red} indicates the best and {\color{blue}blue} indicates the second best. The image size of HR is set to $1280 \times 720$ to calculate the Mult-Adds.}
	\smallskip
	\label{tab:sota}
	\resizebox{0.78\linewidth}{!}{
		\begin{tabular}{|l|c|c|c|c|c|c|c|c|}
			\hline
			\multirow{2}{*}{Method} & \multirow{2}{*}{Scale} & \multirow{2}{*}{Params} & \multirow{2}{*}{Mult-Adds} & Set5 & Set14 & BSD100 & Urban100 & Manga109 \\
			\cline{5-9}
			& & & & PSNR / SSIM & PSNR / SSIM & PSNR / SSIM & PSNR / SSIM & PSNR / SSIM \\
			\hline
			\hline
			Bicubic & \multirow{15}{*}{x2} & - & - & 33.66 / 0.9299 & 30.24 / 0.8688 & 29.56 / 0.8431 & 26.88 / 0.8403 & 30.80 / 0.9339 \\
			SRCNN~\cite{SRCNN} & & 8K & 52.7G & 36.66 / 0.9542 & 32.45 / 0.9067 & 31.36 / 0.8879 & 29.50 / 0.8946 & 35.60 / 0.9663 \\
			FSRCNN~\cite{FSRCNN} & & 13K & 6.0G & 37.00 / 0.9558 & 32.63 / 0.9088 & 31.53 / 0.8920 & 29.88 / 0.9020 & 36.67 / 0.9710 \\
			VDSR~\cite{VDSR} & & 666K & 612.6G & 37.53 / 0.9587 & 33.03 / 0.9124 & 31.90 / 0.8960 & 30.76 / 0.9140 & 37.22 / 0.9750 \\
			DRCN~\cite{DRCN} & & 1774K & 9,788.7G & 37.63 / 0.9588 & 33.04 / 0.9118 & 31.85 / 0.8942 & 30.75 / 0.9133 & 37.55 / 0.9732 \\
			LapSRN~\cite{LapSRN} & & 253K & 104.9G & 37.52 / 0.9591 & 32.99 / 0.9124 & 31.80 / 0.8952 & 30.41 / 0.9103 & 37.27 / 0.9740 \\
			DRRN~\cite{DRRN} & & 298K & 6,796.9G & 37.74 / 0.9591 & 33.23 / 0.9136 & 32.05 / 0.8973 & 31.23 / 0.9188 & 37.88 / 0.9749 \\
			MemNet~\cite{MemNet} & & 678K & 623.9G & 37.78 / 0.9597 & 33.28 / 0.9142 & 32.08 / 0.8978 & 31.31 / 0.9195 & 37.72 / 0.9740 \\
			IDN~\cite{IDN} & & 579K & 133.0G & 37.83 / 0.9600 & 33.30 / 0.9148 & 32.08 / 0.8985 & 31.27 / 0.9196 & 38.01 / 0.9749 \\
			EDSR-baseline~\cite{EDSR} & & 1370K & 316.2G & 37.91 / 0.9602 & 33.53 / 0.9172 & 32.15 / 0.8995 & 31.99 / 0.9270 & 38.40 / 0.9766 \\
			CARN~\cite{CARN} & & 1592K & 222.8G & 37.76 / 0.9590 & 33.52 / 0.9166 & 32.09 / 0.8978 & 31.92 / 0.9256 & 38.36 / 0.9765 \\
			IMDN~\cite{IMDN} & & 694K & 158.8G & 38.00 / 0.9605 & 33.63 / 0.9177 & 32.19 / 0.8996 & 32.17 / 0.9283 & {\color{blue}38.88} / {\color{blue}0.9774} \\
			LatticeNet~\cite{LatticeNet} & & 756K & 169.5G & {\color{red}38.15} / {\color{red}0.9610} & {\color{red}33.78} / {\color{red}0.9193} & {\color{blue}32.25} / {\color{blue}0.9005} & {\color{red}32.43} / {\color{blue}0.9302} & - / - \\
			RFDN~\cite{RFDN} & & 534K & 123.0G & 38.05 / 0.9606 & 33.68 / 0.9184 & 32.16 / 0.8994 & 32.12 / 0.9278 & {\color{blue}38.88} / 0.9773\\
			\textbf{FMEN} & & 748K & 172.0G & {\color{blue}38.10} / {\color{blue}0.9609} & {\color{blue}33.75} / {\color{blue}0.9192} & {\color{red}32.26} / {\color{red}0.9007} & {\color{blue}32.41} / {\color{red}0.9311} & {\color{red}38.95} / {\color{red}0.9778} \\
			\hline
			\hline
			
			Bicubic & \multirow{15}{*}{x3} & - & - & 30.39 / 0.8682 & 27.55 / 0.7742 & 27.21 / 0.7385 & 24.46 / 0.7349 & 26.95 / 0.8556 \\
			SRCNN~\cite{SRCNN} & & 8K & 52.7G & 32.75 / 0.9090 & 29.30 / 0.8215 & 28.41 / 0.7863 & 26.24 / 0.7989 & 30.48 / 0.9117 \\
			FSRCNN~\cite{FSRCNN} & & 13K & 5.0G & 33.18 / 0.9140 & 29.37 / 0.8240 & 28.53 / 0.7910 & 26.43 / 0.8080 & 31.10 / 0.9210 \\
			VDSR~\cite{VDSR} & & 666K & 612.6G & 33.66 / 0.9213 & 29.77 / 0.8314 & 28.82 / 0.7976 & 27.14 / 0.8279 & 32.01 / 0.9340 \\
			DRCN~\cite{DRCN} & & 1774K & 9,788.7G & 33.82 / 0.9226 & 29.76 / 0.8311 & 28.80 / 0.7963 & 27.15 / 0.8276 & 32.24 / 0.9343 \\
			LapSRN~\cite{LapSRN} & & 290K & 115.0G & 33.81 / 0.9220 & 29.79 / 0.8325 & 28.82 / 0.7980 & 27.07 / 0.8275 & 32.21 / 0.9350 \\
			DRRN~\cite{DRRN} & & 298K & 6,796.9G & 34.03 / 0.9244 & 29.96 / 0.8349 & 28.95 / 0.8004 & 27.53 / 0.8378 & 32.71 / 0.9379 \\
			MemNet~\cite{MemNet} & & 678K & 623.9G & 34.09 / 0.9248 & 30.00 / 0.8350 & 28.96 / 0.8001 & 27.56 / 0.8376 & 32.51 / 0.9369 \\
			IDN~\cite{IDN} & & 588K & 60.1G & 34.14 / 0.9259 & 30.13 / 0.8383 & 28.98 / 0.8026 & 27.86 / 0.8463 & 33.11 / 0.9416 \\
			EDSR-baseline~\cite{EDSR} & & 1554K & 160.4G & 34.28 / 0.9263 & 30.24 / 0.8405 & 29.06 / 0.8044 & 28.00 / 0.8493 & 33.37 / 0.9432 \\
			CARN~\cite{CARN} & & 1592K & 118.9G & 34.29 / 0.9255 & 30.29 / 0.8407 & 29.06 / 0.8034 & 28.06 / 0.8493 & 33.50 / 0.9440\\
			IMDN~\cite{IMDN} & & 703K & 71.5G & 34.36 / 0.9270 & 30.32 / 0.8417 & 29.09 / 0.8046 & 28.17 / 0.8519 & 33.61 / 0.9445\\
			LatticeNet~\cite{LatticeNet} & & 765K & 76.3G & {\color{red}34.53} / {\color{red}0.9281} & {\color{blue}30.39} / {\color{blue}0.8424} & {\color{blue}29.15} / {\color{blue}0.8059} & {\color{red}28.33} / {\color{blue}0.8538} & - / - \\
			RFDN~\cite{RFDN} & & 541K & 55.4G & 34.41 / 0.9273 & 30.34 / 0.8420 & 29.09 / 0.8050 & {\color{blue}28.21} / 0.8525 & {\color{blue}33.67} / {\color{blue}0.9449}\\
			\textbf{FMEN} & & 757K & 77.2G & {\color{blue}34.45} / {\color{blue}0.9275} & {\color{red}30.40} / {\color{red}0.8435} & {\color{red}29.17} / {\color{red}0.8063} & {\color{red}28.33} / {\color{red}0.8562} & {\color{red}33.86} / {\color{red}0.9462} \\
			\hline
			\hline
			
			Bicubic & \multirow{15}{*}{x4} & - & - & 28.42 / 0.8104 & 26.00 / 0.7027 & 25.96 / 0.6675 & 23.14 / 0.6577 & 24.89 / 0.7866 \\
			SRCNN~\cite{SRCNN} & & 8K & 52.7G & 30.48 / 0.8626 & 27.50 / 0.7513 & 26.90 / 0.7101 & 24.52 / 0.7221 & 27.58 / 0.8555 \\
			FSRCNN~\cite{FSRCNN} & & 13K & 4.6G & 30.72 / 0.8660 & 27.61 / 0.7550 & 26.98 / 0.7150 & 24.62 / 0.7280 & 27.90 / 0.8610 \\
			VDSR~\cite{VDSR} & & 666K & 612.6G & 31.35 / 0.8838 & 28.01 / 0.7674 & 27.29 / 0.7251 & 25.18 / 0.7524 & 28.83 / 0.8870 \\
			DRCN~\cite{DRCN} & & 1774K & 9,788.7G & 31.53 / 0.8854 & 28.02 / 0.7670 & 27.23 / 0.7233 & 25.14 / 0.7510 & 28.93 / 0.8854 \\
			LapSRN~\cite{LapSRN} & & 543K & 139.3G & 31.54 / 0.8852 & 28.09 / 0.7700 & 27.32 / 0.7275 & 25.21 / 0.7562 & 29.09 / 0.8900 \\
			DRRN~\cite{DRRN} & & 298K & 6,796.9G & 31.68 / 0.8888 & 28.21 / 0.7720 & 27.38 / 0.7284 & 25.44 / 0.7638 & 29.45 / 0.8946 \\
			MemNet~\cite{MemNet} & & 678K & 623.9G & 31.74 / 0.8893 & 28.26 / 0.7723 & 27.40 / 0.7281 & 25.50 / 0.7630 & 29.42 / 0.8942 \\
			IDN~\cite{IDN}  & & 600K & 34.5G & 31.93 / 0.8923 & 28.45 / 0.7781 & 27.48 / 0.7326 & 25.81 / 0.7766 & 30.04 / 0.9026 \\ 
			EDSR-baseline~\cite{EDSR} & & 1518K & 114.2G & 31.98 / 0.8927 & 28.55 / 0.7805 & 27.54 / 0.7348 & 25.90 / 0.7809 & 30.24 / 0.9053 \\
			CARN~\cite{CARN} & & 1592K & 90.9G & 32.13 / 0.8937 & 28.60 / 0.7806 & 27.58 / 0.7349 & 26.07 / 0.7837 & 30.47 / 0.9084 \\
			IMDN~\cite{IMDN} & & 715K & 40.9G & 32.21 / 0.8948 & 28.58 / 0.7811 & 27.56 / 0.7353 & 26.04 / 0.7838 & 30.45 / 0.9075 \\
			LatticeNet~\cite{LatticeNet} & & 777K & 43.6G & {\color{red}32.30} / {\color{red}0.8962} & {\color{blue}28.68} / {\color{blue}0.7830} & {\color{blue}27.62} / {\color{blue}0.7367} & {\color{blue}26.25} / {\color{blue}0.7873} & - / - \\
			RFDN~\cite{RFDN} & & 550K & 31.6G & {\color{blue}32.24} / 0.8952 & 28.61 / 0.7819 & 27.57 / 0.7360 & 26.11 / 0.7858 & {\color{blue}30.58} / {\color{blue}0.9089}\\
			\textbf{FMEN} & & 769K & 44.2G & {\color{blue}32.24} / {\color{blue}0.8955} & {\color{red}28.70} / {\color{red}0.7839} & {\color{red}27.63} / {\color{red}0.7379} & {\color{red}26.28} / {\color{red}0.7908} & {\color{red}30.70} / {\color{red}0.9107} \\
			\hline
	\end{tabular}}
\end{table*}

\begin{figure*}[htbp]
	\begin{center}
	    \begin{tabular}{@{}c@{}c@{}}
		    \includegraphics[width=0.40\linewidth]{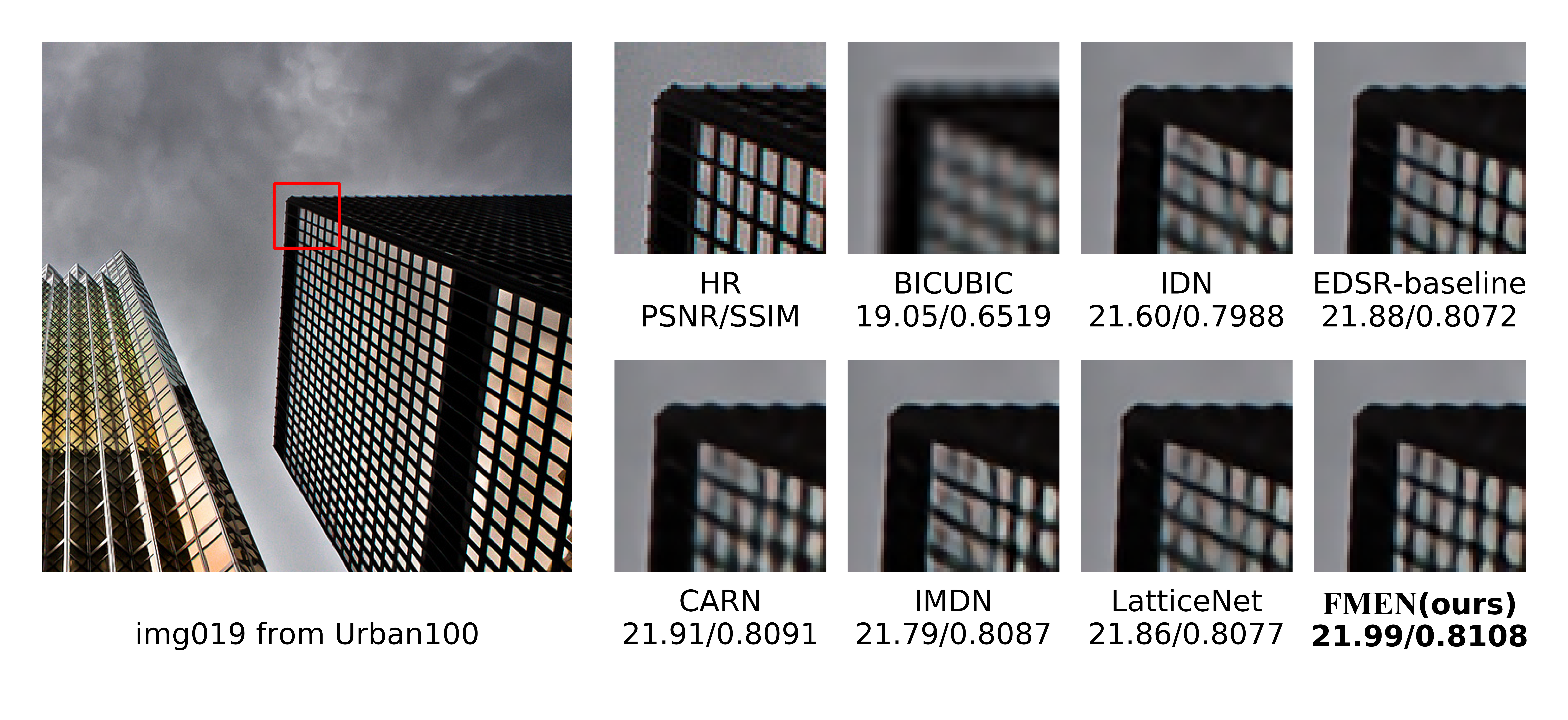} & \includegraphics[width=0.40\linewidth]{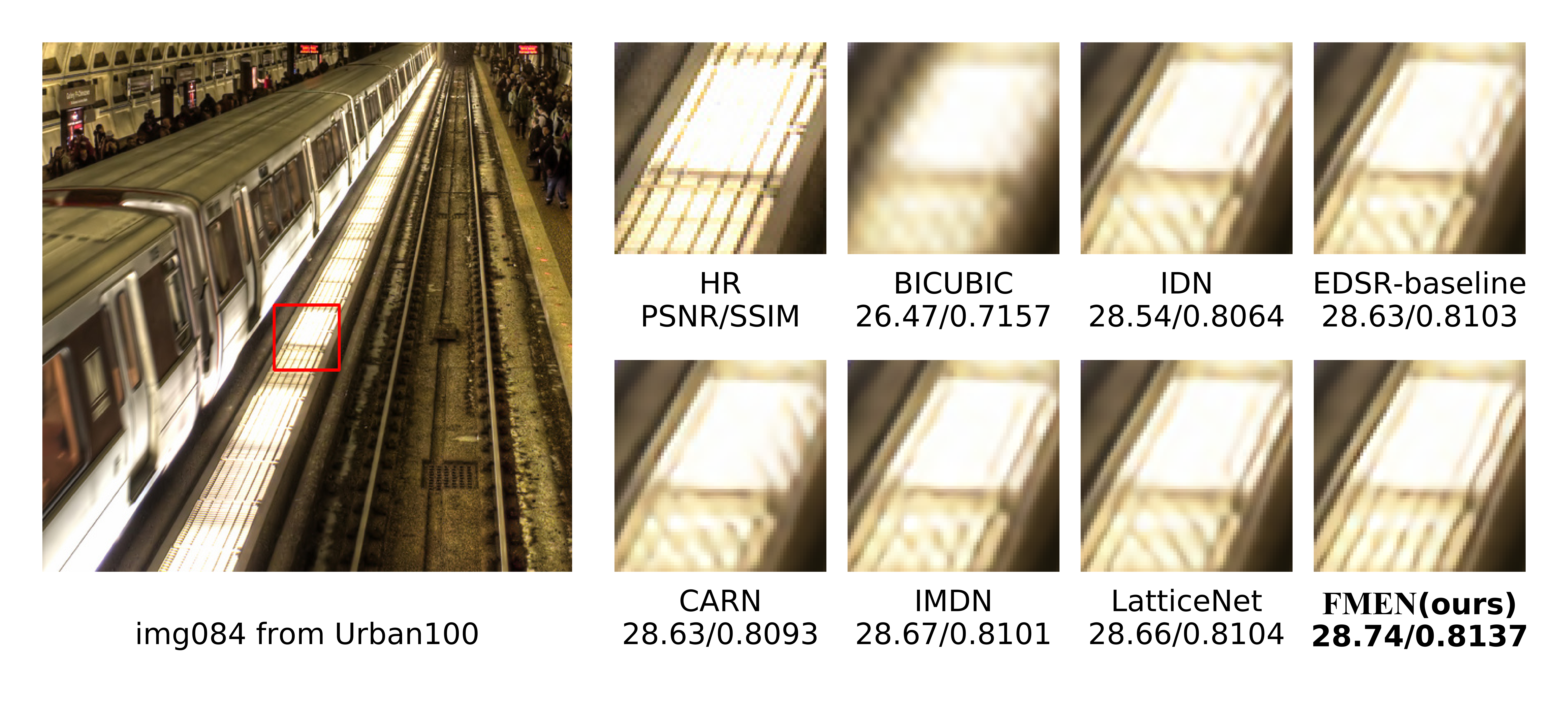}\\
		    \includegraphics[width=0.39\linewidth]{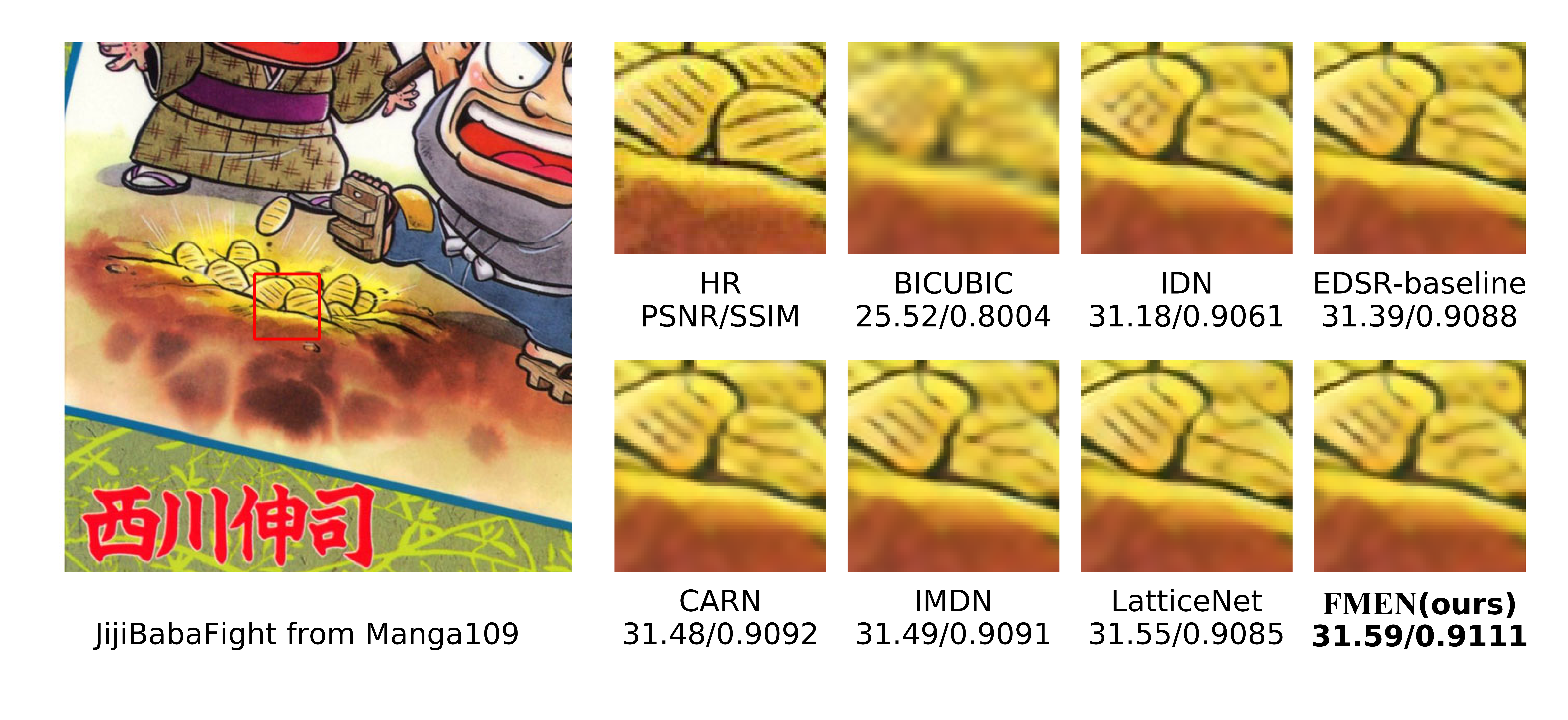} & \includegraphics[width=0.39\linewidth]{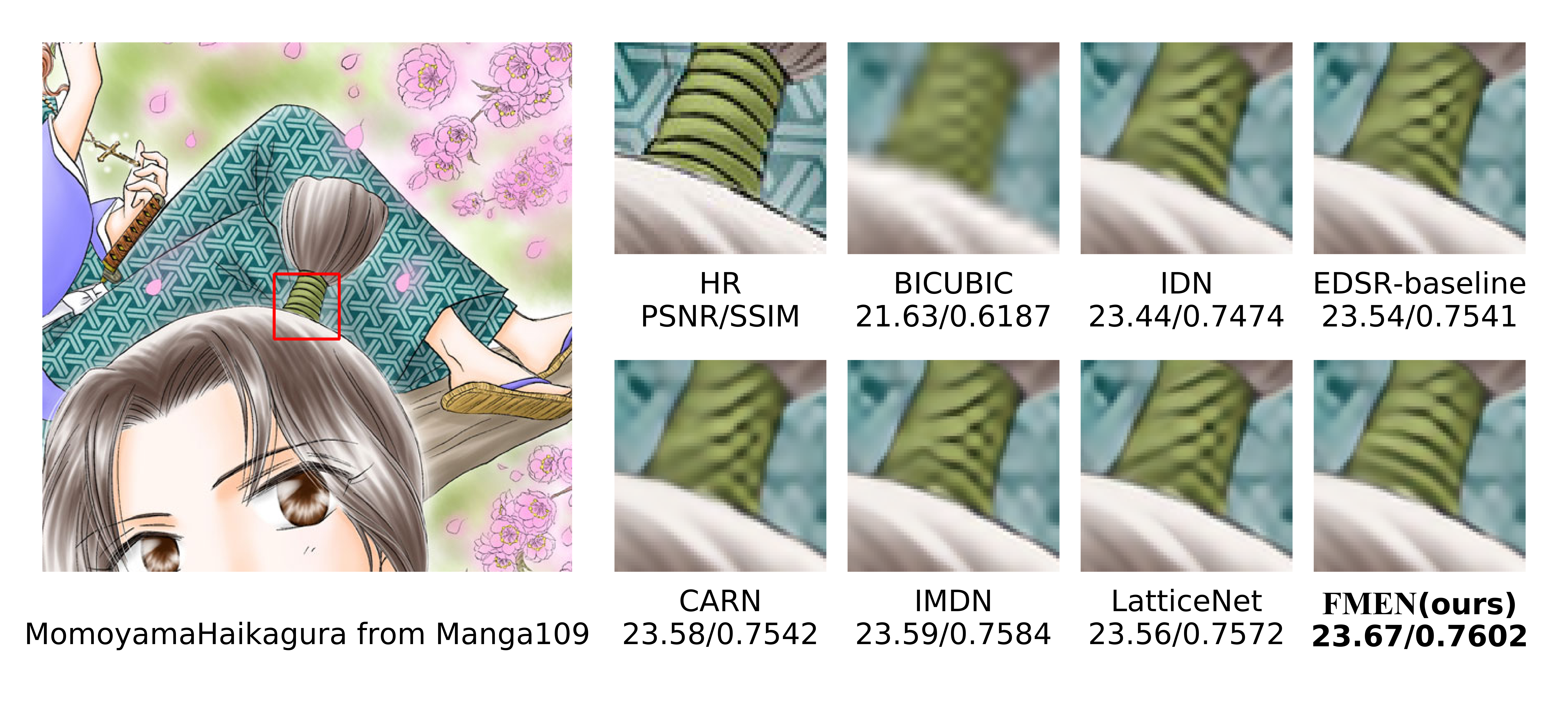}\\
	    \end{tabular}		
	\end{center}
	\vspace{-0.9cm}
	\caption{Visual results on Urban100 and Manga109 for $\times 4$ upscaling.}
	\label{fig:visual}	
\end{figure*}


\subsection{Frequency Analysis}
To figure out whether HFAB focuses on high-frequency areas as we expected, we analyze the feature frequency distribution before and after each HFAB. Inspired by MemNet~\cite{MemNet} and SRFBN~\cite{SRFBN}, we center the power spectra of the average feature maps and estimate spectral densities for a continuous set of frequency by placing concentric circles. We divide spectral density of each frequency by the sum of spectral densities to intuitively observe what percentage low frequency and high frequency account for. The process of calculating power spectrum density (PSD) is displayed in Fig.~\ref{fig:psd}. Based on it, we show the frequency analysis results in Fig.~\ref{fig:frequency}, from which we have three main observations. First, processed by HFAB, low-frequency signals of input features are suppressed and high-frequency signals are strengthened. Second, the inputs (processed by ERBs in the trunk branch) of HFAB slowly focus on high-frequency signals while our HFAB can immediately recalibrate interest areas, which explains why our design has superiority over the stacked RB architecture. Third, the separability of frequency can be further enhanced by posterior HFAB (see the rightmost figure in Fig.~\ref{fig:frequency}).

More intuitively, we plot the inputs and outputs of the four fine grained HFABs (The first HFAB seems to adjust middle-frequency signals in Fig.~\ref{fig:frequency}). The visualization shown in Fig.~\ref{fig:feature} is an efficient complement to the conclusions drawn from Fig.~\ref{fig:frequency}, which provides a deeper insight that the enhanced areas are almost edges and other details.
\begin{table*}[htbp]
	\centering
	\caption{Efficiency comparison on DIV2K validation set for x4 upscaling, with PyTorch 1.4.0, CUDA Toolkit 10.0.130, cuDNN 7.6.5, on an NVIDIA 1080Ti GPU. {\color{red}Red} indicates the best and {\color{blue}blue} indicates the second best. }
	\smallskip
	\label{tab:metrics}
	\resizebox{0.7\linewidth}{!}{
		\begin{tabular}{|l|c|c|c|c|c|c|c|c|}
			\hline
			Method & Params & FLOPs & Runtime & Memory & Activations & Convs & DIV2K val.\\
			\hline
			\hline
			IMDN~\cite{IMDN} & 894K & 58.53G & 50.86ms & 471.76M & 154.14M & 43 & 29.13\\
			\hline
			E-RFDN~\cite{RFDN} & 433K & 27.1G & 41.97ms & 788.13M & 112.03M & 64 & 29.04\\
			\hline
			ByteESR & 317K & 19.7G & {\color{red}27.11ms} & 377.91M & 80.05M & {\color{blue}39} & 29.00\\	\hline
			NEESR & {\color{red}272K} & {\color{red}16.86G} & 29.97ms & 575.99M & {\color{blue}79.59M} & 59 & 29.01\\
            \hline
			Super & 326K & 20.06G & 32.09ms & 663.07M & 93.82M & 59 & 29.00\\
            \hline
			MegSR & 290K & 17.7G & 32.59ms & 640.63M & 91.72M & 64 & 29.00\\
            \hline
			rainbow & {\color{blue}276K} & {\color{blue}17.98G} & 34.1ms & {\color{blue}309.23M} & 92.8M & 59 & 29.01\\	
			\hline
			\textbf{FMEN-S(ours)} & 341K & 22.28G & {\color{blue}28.07ms} & {\color{red}204.6M} & {\color{red}72.09M} & {\color{red}34} & 29.00\\
			\hline
	\end{tabular}}
\end{table*}

\subsection{Comparison with Other Attention Mechanisms}
\label{ex:attention}
%
As we all know, the attention mechanism has been shown useful to boost the performance of EISR networks.  
In this subsection, we compare recent attention mechanisms with our scheme. As discussed in ~\cite{IMDN}, depth is most related to the execution speed, so we construct a baseline with 15 ERBs in the detail learning part, which has the similar depth and number of parameters as ours. Since there are 5 HFABs in our network, we insert an attention block every 3 ERBs into the baseline so that attention mechanism is applied 5 times as well.
Specifically, we investigate contrast-aware channel attention (CCA)~\cite{IMDN, LatticeNet}, and enhanced spatial attention (ESA)~\cite{RFA}. The inserted mode keeps the same with FMEN. Although high learning rate is more beneficial for FMEN, we set the learning rate to $1\times 10^{-4}$ to suit for other attention models. Table.~\ref{tab:attention} shows that the performance of baseline can be improved by embedding current attention blocks, but the gain is relatively lower than that of our scheme. Moreover, with the same number of attention blocks, competing attention mechanisms cause larger time overhead due to MAC brought by multi-branch topology~\cite{IMDN, LatticeNet} and inefficient operations such as $7\times 7$ convolution~\cite{RFDN}, while our scheme is more friendly for network inference without time-consuming operations.

\subsection{Comparison with State-of-the-art Methods}
In this section, we compare our method with other lightweight SR models: SRCNN~\cite{SRCNN}, FSRCNN~\cite{FSRCNN}, VDSR~\cite{VDSR}, DRCN~\cite{DRCN}, LapSRN~\cite{LapSRN}, DRRN~\cite{DRRN}, MemNet~\cite{MemNet}, IDN~\cite{IDN}, EDSR-baseline~\cite{EDSR}, CARN~\cite{CARN}, IMDN~\cite{IMDN}, LatticeNet~\cite{LatticeNet} and RFDN~\cite{RFDN}. 
The quantitative comparisons for x2, x3, x4 upscaling on five publicly available SR benchmark datasets are shown in Tab.~\ref{tab:sota}. FMEN achieves comparable PSNR and SSIM with the most competitive EISR method, LatticeNet, but gains clear advantage in terms of runtime (46ms vs. 68ms) and memory consumption (68M vs. 225M). Visual comparisons are illustrated in Fig.~\ref{fig:visual}. Our method yields more visually pleasant patterns in the selected Urban100 and Manga109 images, compared with other state-of-the-art methods.

\subsection{NTIRE 2022 Challenge on Efficient Super-Resolution}
To participate in this competition,
we reduce the number of convolution kernels to 50 in every convolution layer except for HFAB to maintain the PSNR of 29.00dB on DIV2K validation set. We denote this model as \textit{FMEN-S}. It is worth mentioning that the baseline of the competition~\cite{aim2022}, RFDN, reduces the number of RFDB from 6 to 4 and the number of parameters decreases from 550K to 433K, and it is denoted as \textit{E-RFDN} here~\cite{RFDN}.
We follow the official evaluation setting and report the number of parameters, FLOPs, runtime, peak memory consumption, activations and number of convolution in Tab.~\ref{tab:metrics}. 
Recent advanced EISR methods IMDN~\cite{IMDN}, E-RFDN~\cite{RFDN} and the top methods in~\cite{aim2022} are included for comparison.
Our method achieves the lowest memory consumption and the second shortest runtime, while maintaining comparable restoration accuracy.
Specifically, compared with AIM 2020 winner solution E-RFDN~\cite{RFDN}, our model can decrease 21.2\% parameters, 17.8\% FLOPs, 33.1\% runtime, 74\% peak memory consumption and 35.7\% activations, with only 0.04dB PSNR drop. Compared with other participants in NTIRE 2022 challenge on efficient super-resolution~\cite{aim2022}, our model achieves the best memory consumption, number of activations and convolutions, and the second best inference speed. In contrast to the theoretical metrics, the above metrics are more important in the practical use.

\section{Conclusions}
In this paper, we analyze the factors which influence runtime and memory consumption of current EISR models, and design a fast and memory-efficient network (FMEN) with efficient sequential operators and attention mechanism. FMEN is mainly composed of two basic blocks: enhanced residual block (ERB) and high-frequency attention block (HFAB). ERB takes advantage of RB but more friendly for deployment. HFAB is more powerful and lightweight than conventional attention blocks. Our method gains substantial improvement of runtime and memory consumption, while maintaining comparable reconstruction performance.

\clearpage
\newpage
{\small
\bibliographystyle{ieee_fullname}
\bibliography{egbib}
}

\end{document}


\title{High Frequency Attention Network for Efficient Image Super-Resolution\\(Supplementary Material)}

\author{Zongcai Du \quad Ding Liu
Institution1\\
Institution1 address\\
{\tt\small firstauthor@i1.org}
\and
Second Author\\
Institution2\\
First line of institution2 address\\
{\tt\small secondauthor@i2.org}
}
\maketitle
\begin{table}
    \centering
    \caption{The number of long skip connection and group convolution which cause significant MAC.}
    \resizebox{1.0\linewidth}{!}{
    \begin{tabular}{cccccc}
    \toprule
    \diagbox{Operation}{Model} & CARN~\cite{CARN} & IMDN~\cite{IMDN} & RFDN~\cite{RFDN} & LatticeNet~\cite{LatticeNet} & HFAN\\
    \hline
    Addition & 9 & 13 & 17 & 33 & 1\\
    \hline
    Multiplication & - & 6 & 4 & 16 & 5\\
    \hline
    Concatenation & 12 & 7 & 5 & 10 & -\\
    \hline
    Group convolution & 12 & - & - & - & 5\\
    \hline
    Total & 33 & 26 & 26 & 59 & \textbf{11} \\
    \bottomrule
    \end{tabular}}
    \label{tab:mac}
\end{table}

\begin{table}
    \centering
    \caption{The number of extra $1\times 1$ convolution for local and global feature fusion.}
    \resizebox{1.0\linewidth}{!}{
    \begin{tabular}{cccccc}
    \toprule
    $1\times 1$ convolution & \multirow{2}{*}{CARN~\cite{CARN}} & \multirow{2}{*}{IMDN~\cite{IMDN}} & \multirow{2}{*}{RFDN~\cite{RFDN}} & \multirow{2}{*}{LatticeNet~\cite{LatticeNet}} & \multirow{2}{*}{HFAN}\\
    input channels $\rightarrow$ output channels & & & & & &
    \hline
    48 $\rightarrow$ 24 & - & - & 12 & - & -\\
    \hline
    64 $\rightarrow$ 32 & - & - & - & 6 & -\\
    \hline
    64 $\rightarrow$ 64 & - & 6 & - & - & -\\
    \hline
    96 $\rightarrow$ 48 & - & - & 4 & - & -\\
    \hline
    128 $\rightarrow$ 64 & 4 & - & - & - & -\\
    \hline
    192 $\rightarrow$ 64 & 4 & - & - & - & -\\
    \hline
    256 $\rightarrow$ 64 & 4 & - & 1 & - & -\\
    \hline
    384 $\rightarrow$ 64 & - & 1 & - & - & -\\
    \hline
    Total & 12 & 7 & 17 & 6 & \textbf{0} \\
    \bottomrule
    \end{tabular}}
    \label{tab:1x1}
\end{table}

\begin{table}[t]
    \centering
    \caption{Efficiency comparison between different type of $1\times 1$ convolution and normal $3\times 3$ convolution with stride=1, cuDNN 7.6.5 on an NVIDIA 1080Ti. The runtime is an average of 10 runs after warming up the hardware. We set the input size to $421\times 421$ to calculate FLOPs, which is the same as the average image size of validation set.}
    \resizebox{1.0\linewidth}{!}{
    \begin{tabular}{cccc}
    \toprule
    $1\times 1$ convolution & \multirow{2}{*}{FLOPs(G)} & \multirow{2}{*}{Runtime(ms)} & \multirow{2}{*}{Ratio}\\
    input channels $\rightarrow$ output channels & & & &
    \hline
    48 $\rightarrow$ 24 & 0.204 & 0.286 & 0.713\\
    \hline
    64 $\rightarrow$ 32 & 0.363 & 0.378 & 0.960\\
    \hline
    64 $\rightarrow$ 64 & 0.726 & 0.598 & 1.214\\
    \hline
    96 $\rightarrow$ 48 & 0.817 & 0.947 & 0.863\\
    \hline
    128 $\rightarrow$ 64 & 1.452 & 1.261 & 1.152\\
    \hline
    192 $\rightarrow$ 64 & 2.178 & 1.745 & 1.248\\
    \hline
    256 $\rightarrow$ 64 & 2.904 & 2.154 & 1.348\\
    \hline
    384 $\rightarrow$ 64 & 4.356 & 3.022 & 1.441\\
    \hline
    \hline
    $3\times 3$ convolution & \multirow{2}{*}{6.534} & \multirow{2}{*}{1.298} & \multirow{2}{*}{5.034}\\
    64 $\rightarrow$ 64 & & & &
    \bottomrule
    \end{tabular}}
    \label{tab:time}
\end{table}
\section{Inference Speed Analysis}
%
Our model runs much faster than recent efficient image super-resolution (EISR) models, such as CARN~\cite{CARN}, IMDN~\cite{IMDN}, LatticeNet~\cite{LatticeNet} and RFDN~\cite{RFDN}, because of reducing memory access cost (MAC) and avoiding feature fusion scheme which brings about multiple extra $1\times 1$ convolutions.
In terms of MAC, it consists of a large portion of time usage in long skip connection (addition, subtraction, multiplication, division and concatenation) and group-wise convolution~\cite{RepVGG}.
We show the number of these operations of recent models in Tab.~\ref{tab:mac}. Our model reduces more than half amount of memory access time compared with RFDN.
For extra $1\times 1$ convolution used in feature fusion, we first count its type and number according to the input and output channels which is displayed in Tab.~\ref{tab:1x1} and test the inference time of every type. We construct a baseline model with ten $3\times 3$ convolution in the detail learning part and calculate its average inference time $t_{base}$ over DIV2K validation set for x4 upscaling. After that, we repeat the same operator ten times at the end of the detail learning part in the baseline model and calculate its average runtime $t_{new}$. Then the inference time of every convolution is obtained via $\frac{t_{new}-t_{base}}{10}$. Besides, followed RepVGG~\cite{RepVGG}, we calculate the FLOPs of every operator and denote the ratio of FLOPs to runtime as a measurement of its efficiency. The results are shown in Tab.~\ref{tab:time}, from which we can see that $3\times 3$ convolution is much more efficient than $1\times 1$ convolution, and thus more suitable for EISR. By cutting down the usage of long skip connection and avoiding feature fusion, our model can achieve the highest speed (28.1ms) compared with that of RFDN (43.2ms), the winner solution in AIM20~\cite{zhang2020aim}.

{\small
\bibliographystyle{ieee_fullname}
\bibliography{egbib}
}